\documentclass[letterpaper]{article} 
\usepackage{aaai25}  
\usepackage{times}  
\usepackage{helvet}  
\usepackage{courier}  
\usepackage[hyphens]{url}  
\usepackage{graphicx} 
\usepackage{natbib}  
\usepackage{caption} 
\usepackage{algorithm}
\usepackage{algorithmic}

\usepackage{newfloat}
\usepackage{listings}
\DeclareCaptionStyle{ruled}{labelfont=normalfont,labelsep=colon,strut=off} 
\floatstyle{ruled}
\newfloat{listing}{tb}{lst}{}
\floatname{listing}{Listing}
\usepackage{algorithm}
\usepackage{algorithmic}
\usepackage{array}

\usepackage{newfloat}
\usepackage{listings}

\floatstyle{ruled}
\newfloat{listing}{tb}{lst}{}
\floatname{listing}{Listing}

\usepackage{xcolor}
\newcommand{\answerYes}[1]{\textcolor{blue}{#1}} 
\newcommand{\answerNo}[1]{\textcolor{teal}{#1}} 
\newcommand{\answerNA}[1]{\textcolor{gray}{#1}}

\title{Global Patterns of Viral Content on WhatsApp}

\author{
Kiran Garimella\textsuperscript{\rm 1}\footnote{Corresponding author},
Princessa Cintaqia\textsuperscript{\rm 2},
Juan Jose Rojas Constain\textsuperscript{\rm 3},
Bharat Nayak\textsuperscript{\rm 4},
Aditya Vashistha\textsuperscript{\rm 5}
\\
}
\affiliations{
    \textsuperscript{\rm 1}Rutgers University,
    \textsuperscript{\rm 2}Boston University,
    \textsuperscript{\rm 3}Universidad de Los Andes,\\
    \textsuperscript{\rm 4}Independent Journalist,
    \textsuperscript{\rm 5}Cornell University\\
     kiran.garimella@rutgers.edu, 
     cintaqia@bu.edu,
     jj.rojasc1@uniandes.edu.co,\\
     bharatnayak34@gmail.com,
     adityav@cornell.edu
}

\begin{document}


\maketitle

\begin{abstract}
This paper explores the nature and spread of viral WhatsApp content among everyday users in three diverse countries: India, Indonesia, and Colombia. By analyzing hundreds of viral messages collected with participants' consent from private WhatsApp groups, we provide one of the first cross-cultural categorizations of viral content on WhatsApp. 
Despite the differences in cultural and geographic settings, our findings reveal striking similarities in the types of groups users engage with and the viral content they receive, particularly in the prevalence of misinformation.
Our comparative analysis shows that viral content often includes political and religious narratives, with misinformation frequently recirculated despite prior debunking by fact-checking organizations. These parallels suggest that closed messaging platforms like WhatsApp facilitate similar patterns of information dissemination across different cultural contexts. 
This work contributes to the broader understanding of global digital communication ecosystems and provides a foundation for future research on information flow and moderation strategies in private messaging platforms.
\end{abstract}


\section{Introduction}

WhatsApp, a cornerstone of global digital communication, profoundly impacts billions of users worldwide. Despite its widespread adoption, its role as a key channel for information dissemination—particularly in diverse contexts across the Global South—remains significantly underexplored. This oversight is particularly concerning given the severe real-world consequences that misinformation on this platform has instigated, such as social unrest and violence~\cite{arun2019whatsapp}. Our research seeks to bridge this gap by analyzing the spread and nature of `viral' content among distinct populations: villagers in central India, university students in Indonesia and a diverse set of users from Colombia. These regions represent a significant portion of WhatsApp's user base and are emblematic of the challenges and opportunities that arise in the Global South's digital ecosystems.

All three countries we consider, while disparate in their socio-economic and cultural fabrics, reveal the dual nature of WhatsApp as a medium for everyday communication and a tool for widespread misinformation. This study, through a mixed-methods approach involving hundreds of viral messages from private WhatsApp groups donated by users, offers critical insights into the types of digital communities formed and the content that defines their information diets. Notably, our findings expose a high prevalence of misinformation intertwined with religious and political content, highlighting a common thread of digital discourse that transcends geographical and cultural boundaries.

The complexity of researching WhatsApp stems from its encrypted, private nature and the absence of public APIs, which poses significant challenges for systematic data collection and ethical concerns regarding user privacy. Despite these obstacles, our methodology captures a broad spectrum of `forwarded many times' messages, allowing us to identify macro trends in viral content dissemination.

Prior studies in this domain have primarily been qualitative, focusing on small samples, which provide deep insights but can not capture macro trends~\cite{varanasi2022accost}. These studies do not fully appreciate the textured layers of personal WhatsApp use, especially concerning how viral messages resonate within different social groups.

Our findings reveal both striking similarities and notable differences in the nature of WhatsApp group activity across India, Indonesia, and Colombia. Personal groups, such as those formed by family and friends, dominate the dataset in all three countries, highlighting the platform's central role in fostering close-knit communication. However, these groups are not always the primary contributors to the viral content we analyzed. Instead, political and activist groups emerge as key drivers of viral content, underscoring their influence in shaping narratives on the platform.

A significant portion of this viral content is misinformation, with prevalence rates of over 25\% in India, 30\% in Indonesia, and 10\% in Colombia. This highlights the platform's role as a conduit for false or misleading information. In India, much of this misinformation is intertwined with hate speech and religious themes, reflecting deep societal and political divides. In contrast, health misinformation and anti-government narratives dominate in Indonesia and Colombia, pointing to regional variations in the types of content that gain traction.

Perhaps most concerning is the near-total absence of fact-checking or misinformation correction within the groups. Across the wide range of groups in our dataset, only a handful of instances involved members actively challenging false claims. This lack of intervention, even in groups with diverse membership, underscores a critical gap in the collective response to misinformation. These findings emphasize the urgent need for strategies to counter the spread of harmful content in closed messaging environments like WhatsApp.

This paper contributes to Social Computing research by detailing how localized digital practices inform broader usage patterns and affect social dynamics. 
Our findings indicate the urgent need for researchers to focus on underrepresented regions, advancing our understanding of how global platforms like WhatsApp can both support and undermine community resilience and social cohesion. By highlighting the nuanced ways in which misinformation spreads and is contested, we advocate for more inclusive digital communication tools that respect and reinforce local cultural practices, thereby supporting more robust democratic processes and social trust.



\section{Related Work}

The rise of social media platforms like WhatsApp has reshaped communication landscapes globally, but nowhere is this transformation more pronounced than in the Global South. In India, WhatsApp has amassed a user base exceeding 500 million, while in Indonesia, the platform engages around 100 million users. In Colombia, WhatsApp is the most popular social network with over 90\% of internet users using the app. This vast penetration highlights WhatsApp's pivotal role in daily communications, significantly influencing social, political, and cultural dynamics \cite{baulch2020introduction}. These countries, characterized by their high population densities and rapidly expanding digital infrastructures, offer unique insights into the digital behaviors that pervade less-studied regions of the world.

Social media’s exponential growth has, however, been shadowed by the spread of digital misinformation, leading to severe outcomes including lynchings, civic unrest, and increased political polarization \cite{anderson2020, arun2019whatsapp}. Prior research has extensively documented the impact of misinformation in urban settings, but there remains a significant void in our understanding of how it permeates and affects the rural populace, whose information diets are predominantly shaped by mobile-first internet access \cite{arora2016bottom}. This study aims to bridge this gap by digging into the nuances of information consumption in multiple contexts -- users from India, Indonesia, and Colombia thereby providing a comparative glimpse into how misinformation spreads across different community structures within the Global South.

Existing literature often focuses on technologically advanced settings, leaving a gap in comprehensive studies that address the integration of technology in everyday life in rural and semi-urban areas of developing countries. Studies like that by \citet{varanasi2022accost} have begun to explore how digital platforms are adapted to local needs and languages, revealing a complex interplay between technology adoption and cultural context. Furthermore, research by \citet{banaji2019whatsapp} underscores the role of digital platforms in forming new social networks that transcend traditional boundaries such as caste and religion, which are otherwise rigid in rural settings. This background sets the stage for our study, which examines WhatsApp's dual role as both a facilitator of connectivity and a conduit for misinformation, considering the platform's deep embedment into the local cultural and social fabrics of India, Indonesia, and Colombia. 


Prior research has highlighted the pervasive role of WhatsApp in misinformation spread, often linking it to broader socio-political dynamics. \citet{varanasi2022accost} examined how rural and urban communities in India interact with misinformation, highlighting the influence of social status, local norms, and community deliberations on information sharing. Similarly, \citet{syam2020} analyzed COVID-19 misinformation in Indonesian WhatsApp groups, finding that content decisions were driven more by personal beliefs and emotions than by fact-checking, contributing to public panic and diminished trust in the platform. These studies underscore a recurring theme: misinformation is often propagated within trusted social circles, where the identity of the sender outweighs the credibility of the message.

Both contexts show that misinformation is not just a byproduct of digital interaction but is deeply embedded in the socio-political fabric of the societies. For instance, \citet{banaji2019whatsapp} and \citet{arun2019whatsapp} documented the violent repercussions of misinformation in India, particularly in how anti-Muslim sentiments and political propaganda are amplified within closed groups. Comparable patterns were observed in Indonesia during the 2019 presidential elections, where misinformation and `HOAX’ posts were extensively shared due to low digital literacy and emotional engagement with content, as noted by \citet{susilo2020}. 

Across the three countries we study, WhatsApp functions as a platform where political and religious narratives frequently dominate, often exacerbating social divides. \citet{chakrabarti2018duty} identified recurring themes of nationalism, cultural pride, and political superiority in Indian WhatsApp groups, paralleling findings by \citet{neyazi2022}, who observed that Indonesian WhatsApp users were more prone to political misinformation compared to other platforms like Facebook and Twitter. The end-to-end encryption of WhatsApp, while protecting privacy, simultaneously creates challenges for monitoring and mitigating the spread of false information, as \citet{shahid_one_2024} reports.

Despite these shared challenges, a common gap in existing research is the lack of systematic analysis of viral content within private WhatsApp groups at scale, particularly across different types of groups. Our study addresses this gap by examining viral content in hundreds of private groups in rural India, among Indonesian university students, and a diverse set of users in Colombia. By analyzing hundreds of messages, our research extends the current literature by providing empirical evidence of the types of content that gain virality and the group dynamics that facilitate this spread. Our mixed-methods approach offers a nuanced view of how misinformation is not just a consequence of digital media but is also a reflection of underlying social structures, biases, and emotional triggers.

\section{Methodology}

\subsection{Data Collection}
We used a tool built by \citet{garimella2024whatsapp} to collect data from three diverse regions: a village in Jharkhand, India; universities across Indonesia; and a diverse set of users in Colombia. The tool was specifically designed to ensure the privacy and anonymity of participants’ data, featuring a user-friendly interface that allowed participants to donate their WhatsApp data.

The data collection protocols, facilitated by the data donation tool, were well documented and received IRB approval. This ensured that all ethical considerations—particularly around participant privacy and data security—were rigorously upheld. 
For example, the tool filtered out groups with fewer than six members and automatically anonymized personal information, including names, phone numbers, and emails, using Google’s Data Loss Prevention library. Furthermore, any images collected had faces blurred to maintain privacy. Once the data was donated, the tool automatically selected viral messages, i.e.,  content classified as `forwarded many times' on WhatsApp, a label used to signify viral spread within the platform. This designation indicates that a message has been forwarded through a chain of five or more separate users from the original sender, indicating that a large number of people got exposed to this message.\footnote{The exact definition by WhatsApp can be found here \url{https://faq.whatsapp.com/1053543185312573}.}



In India, we focused on a rural village in Jharkhand, an area noted for its communal sensitivity and frequent incidents of mob violence, which emphasizes the importance of studying the spread of misinformation. The recruitment process was conducted on the ground through a local researcher, which facilitated a deeper understanding and trust with participants. A total of 31 male participants from the village donated data, providing insights from 164 WhatsApp groups over two months in June--August 2023. The dataset contained 604 viral messages, from
predominantly 
large groups with a median size of 33 members, ranging from community news to religious and political discussions.

In Indonesia, we targeted a tech-savvy demographic of university students from various universities. Recruitment was done digitally via a popular Twitter account (@collegemenfess) with over 100,000 followers, many of whom are Indonesian students. 
We recruited 74 participants who contributed 177 viral messages from 97 WhatsApp groups which had a median group size of 35. The Indonesian dataset was collected during March--July 2024 and contained information on student life and interactions across different social and religious groups, including groups related to family/friends, work/study, and political activism.

In Colombia, participants were contacted through a database of over 5,000 individuals interested in participating in research studies, facilitated by the E-SocialS research group at the University of Valle, in Colombia. The recruitment and data donation process took place between February and August 2024. A total of 75 participants contributed 610 viral messages from 140 WhatsApp groups. The participants were geographically diverse, with some residing in Cali and nearby villages in the southern region, others in Bogotá, the capital located in the central region, and a portion in villages of the northern region.  

Table~\ref{tab:demographics} shows the details of our datasets, including the sample size and the demographics.
All three contexts provided a unique perspective on how WhatsApp is used in varying social settings, revealing both commonalities and differences in digital communication practices, which have not been explored previously.
The dataset containing all the viral content collected in this paper is available online.\footnote{\url{https://doi.org/10.7910/DVN/VOFPK1} Please check the WhatsApp folder.}


\begin{table*}[t]
\centering
\caption{Details of the datasets from India, Indonesia, and Colombia.}
\label{tab:demographics}
\begin{tabular}{l|c|c|c|c|c|c}
\hline
          & \textbf{N}  & \textbf{Age range} & \textbf{Gender}                                                            & \textbf{Religion}     & \textbf{Num. groups} & \textbf{Num. viral messages}                                                \\
\hline
India     & 31 & 21-34     & Male: 100\%                                                       & Hindu: 100\%  & 164        & 604     \\
\hline
Indonesia & 74 & 18-24     & \begin{tabular}[c]{@{}l@{}}Male: 22\%\\ Female: 78\%\end{tabular} & \begin{tabular}[c]{@{}l@{}}Islam: 70\%\\ Christian: 30\%\end{tabular} & 97 & 177\\
\hline
Colombia & 75 & 18-64     & \begin{tabular}[c]{@{}l@{}}Male: 40\%\\ Female: 60\%\end{tabular} & \begin{tabular}[c]{@{}l@{}}Christian: 61\%\\ Nonreligious: 39\%\end{tabular} & 140 & 610\\
\hline
\end{tabular}
\end{table*}

While our datasets provide unique insights into WhatsApp usage within specific communities in India, Indonesia, and Colombia, these datasets are not fully representative of the broader populations in these regions. In India, the dataset primarily reflects the perspectives of Hindu males, with no participation from the Muslim minority, who make up a significant portion of the village’s population. Similarly, in Indonesia, the focus on university students from Java may not encompass the diverse experiences of individuals from other islands or different demographic groups. In Colombia, while the dataset includes a variety of users, the nature of the recruitment database may have led to an over representation of left-leaning and well-educated individuals.
Despite these limitations, our study represents a significant effort to capture data from underrepresented communities using ethically responsible methods and explicit participant consent. Furthermore, the content analyzed in this study spread virally on WhatsApp, indicating widespread forwarding and engagement. Regardless of the sampling frame, these findings provide valuable and intriguing insights into the dynamics of viral content dissemination.
%


\noindent\textbf{A note on the ethics of our data collection}.
Studying WhatsApp, particularly in contexts within the Global South, is critically important for understanding contemporary information flows and the propagation of misinformation. As demonstrated by our findings, misinformation is highly prevalent on WhatsApp, highlighting the urgency of producing descriptive analyses to inform public awareness, policy interventions, and platform accountability. Given WhatsApp's widespread use and influence in these regions, insights derived from such research are essential to prompting informed discourse and practical solutions to curb misinformation.

WhatsApp remains significantly understudied, particularly in relation to its distinctive affordances such as end-to-end encryption, which renders traditional data collection methods unfeasible. The encrypted nature of the platform means that even WhatsApp itself does not have access to user communications, further limiting direct analysis. Consequently, ethical research practices must be carefully considered and transparently implemented to respect users' privacy and autonomy.

Our approach, utilizing a consent-driven data donation method, represents the most ethical and viable solution for quantitatively studying WhatsApp content. Participants explicitly consented to donate data exclusively from private groups sharing viral messages. By focusing solely on viral content—messages that have already reached a broad audience within the platform—this method minimizes privacy concerns associated with personal or sensitive communications. Data donation thus balances the imperative for rigorous academic research with the ethical responsibility to protect individual privacy and maintain trust among participants.


\subsection{Data Annotation and Analysis}


The dataset contained 1,391  viral messages, 604 from India, 177 from Indonesia, and 610 from Colombia. We utilized a custom dashboard designed to facilitate the annotation process, enabling annotators to view the context of each message by displaying de-identified adjacent messages (10 before and after the focal message). This setup helped in understanding the interactions surrounding the viral content.

All datasets underwent a rigorous qualitative coding process conducted by native experts who were deeply familiar with the cultural and social contexts of the respective datasets. These annotators were responsible for categorizing each message based on emergent themes and refining these into organized parent categories through multiple iterations of review and discussion among the research team.

In India, the coding was carried out by a skilled fact-checker but also native to the studied village, providing an intrinsic understanding of local dynamics such as caste hierarchies and community-specific linguistics. 
The annotation process employed an inductive coding strategy, allowing codes to emerge organically and evolve through iterative refinement. This approach identified 11 significant categories, including combinations such as misinformation intertwined with hate speech targeting Muslims. Each piece of content was evaluated for its alignment with multiple categories, reflecting the complex and overlapping nature of the themes. Misinformation was particularly scrutinized; each piece of content was checked for the accuracy of its claims using tools like reverse image searches and standard internet searches for text verification.

In Indonesia, the annotator was a native Indonesian and a university student, which was crucial for interpreting the content within the contemporary social and educational contexts of Indonesian youth. The categories derived from the Indian annotations served as a baseline, with additional inductive coding conducted to capture context-specific nuances in the Indonesian dataset. This resulted in 13 distinct categories, including 2 additional categories: advertising and pro-China content. The identification of misinformation was facilitated by the Indonesian Communication Agency's online resources, which often include `HOAX' verification articles that helped in validating the misinformation claims found in viral posts.

In Colombia, the annotation was conducted by a native professional in collaboration with a journalist who leads a prominent Colombian fact-checking unit. Through an iterative and collaborative process, the categorization resulted in 13 specific types of content. The annotated content included diverse types of misinformation, and any suspicious information that lacked a prior fact-check was subjected to thorough verification.

The annotation process in all three countries also included documenting whether the viral content was fact-checkable and if it had been fact-checked. Additionally, all the annotators qualitatively analyzed and recorded the types of responses each viral post received within the group chats, such as corrections or expressions of support. This not only categorized the content but also provided insights into the social reactions to misinformation, enhancing our understanding of how misinformation spreads and is received in different cultural contexts. The detailed code book containing the rubric for the three countries is available in the Appendix.

In addition to annotating the viral content, the types of groups were also annotated based on the name of the group and types of viral messages observed in the groups.


\section{Findings}
We structure our findings as follows. First, we present analysis on the types of groups found in the three contexts. Next, we look at popular narratives across countries looking at similarities and differences across the countries. Finally, we dig specifically into significant types of content -- misinformation, entertainment, and fact checking content.

\begin{table}[]
\caption{Categories of the WhatsApp groups.}
\label{tab:group_labels}
\centering
\resizebox{\columnwidth}{!}{%
\begin{tabular}{cccc}
\hline
\textbf{Category} & \textbf{India \%} & \textbf{Indonesia \%} & \textbf{Colombia \%} \\
\hline
Family & 6.6 & 15.6 & 9.3\\
Friends & 9.7 & 32.8 & 9.3\\
Regional & 15.2 & 17.1 & 11.4\\
Village/Town & 6.7 & 9.3 & -\\
News & 12.3 & 4.6 & -\\
Religious & 6.7 & 25.0 & 1.4\\
Activism & 12.2 & 3.1 & 9.3\\
Politics & 8.5 & - & 8.6 \\
Caste & 7.3 & - & -\\
Study-related & - & 31.2 & 15.0\\
Work-related & - & - & 6.4 \\
Sales & - & - & 6.4 \\
Jobs/Opportunities & - & - & 13.6 \\
Others & 6.6 & 6.2 & 9.3\\
\hline
\end{tabular}
}
\end{table}

\subsection{Types of groups}

Table~\ref{tab:group_labels} presents the categorization of WhatsApp groups analyzed in this study. Across all three countries, regional and personal groups (e.g., family and friends) constituted the largest share of groups.

In India, although caste-based and religious (Hindutva) groups comprised only 13\% of the total groups, they accounted for a substantial proportion of the viral messages (40\%). This disproportionality indicates that these groups are not only highly active but also play a critical role in disseminating viral messages, often laden with misinformation and politically charged narratives.
In Indonesia, religious and study groups were predominant, both in terms of group prevalence and their share of viral content. In contrast, Colombia exhibited a more balanced distribution of groups, with notable categories related to jobs and study groups.

Although explicit political groups were present in India and Colombia but absent in Indonesia, political content was found to permeate non-political groups across all three countries. This underscores the pervasive nature of political discourse on WhatsApp, transcending the boundaries of explicitly political spaces.

\begin{table}[h]
\centering
\caption{Categories of viral content on WhatsApp in India, Indonesia and Colombia. The percentage does not sum up to 100 since one post can be in more than one category.}
\label{tab:combined_categories}
\resizebox{\columnwidth}{!}{%
\begin{tabular}{p{4.3cm}|p{1.2cm}|p{1.5cm}|p{1.5cm}}

\hline
\textbf{Category} & \textbf{India \%} & \textbf{Indonesia \%} & \textbf{Colombia \%} \\ \hline
Misinformation & 26 & 30.5 & 10.5\\ \hline
National/International News & 23.2 & 29.8 & 10.5\\ \hline
Inspirational/Informational Content & 21.8 & 20.9 & 15.4\\ \hline
Religious Propaganda & 21.0 & 4.5 & 0.5 \\ \hline
Hate Speech & 19.6 & 2.8 & 0.5 \\ \hline
Political Propaganda & 18.6 & 3.9 & 23.8 \\ \hline
Regional Information & 8.5 & 32.7 & 21.6 \\ \hline
Religious Content & 7.3 & 7.3 & 2.3 \\ \hline
Entertainment/Humor/Sarcasm & 6.2 & 14.1 & 11.1\\ \hline
Good Morning Messages & 3.5 & 1.1 & 2.9 \\ \hline
Health information & 1.4 & 6.7 & 3.0\\ \hline
Advertising & - & 31.0 & 23.3 \\ \hline

Chinese Content & - & 6.7 & - \\ \hline
\end{tabular}
}
\end{table}
\subsection{Popular viral narratives}

In the milieu of frequently forwarded messages on WhatsApp, certain narratives demonstrated a recurring pattern. These narratives, often entrenched in political and social discourse, not only hold sway over public opinion but also serve to perpetuate existing divides. This section presents the predominant narratives that surfaced in our datasets. Though there is a significant overlap in the categories of content (see Table~\ref{tab:combined_categories}), the differences also make the findings interesting. We will qualitatively discuss the prominent narratives separately in India, Indonesia, and Colombia in the rest of this section.
We will reflect on the similarities and differences overall in Section~\ref{sec:discussion}.

\subsubsection{India:}
\noindent (i) \textbf{PM Modi as a Global and National Leader}. One persistent narrative positioned Prime Minister Narendra Modi as an unparalleled leader, respected both nationally and globally. In this narrative, Modi was often depicted as the `savior' of Hindus---a figure capable of restoring Hindu pride and protecting the community from external threats. Messages with such content often included exaggerated accolades, overlooking any criticisms or complexities related to his tenure. These narratives served to consolidate the Prime Minister's standing and reinforce the party's core voter base and seem to be a persistent narrative being pushed by the BJP, since such narratives were also prevalent in qualitative work by~\citet{chakrabarti2018duty} almost 7 years ago.

\noindent (ii) \textbf{Fear-Mongering and Communal Tensions}. A more concerning narrative revolved around instigating fear among Hindus. We identified a pervasive climate of distrust and antipathy towards Muslims, equating them to traitors who belong to Pakistan. Messages in this category implied that failing to vote for the BJP would lead to an impending demographic shift where Muslims become the majority, thereby posing a threat to Hindu safety. Hindu women were often specifically portrayed as victims in such messages, magnifying the perceived threat. This narrative also extended to the vilification of particular minority groups, such as Rohingya Muslims and Bangladeshi immigrants, as disruptive elements. Calls to boycott Muslim shopkeepers and their products were common, along with the propagation of conspiracy theories like `Love Jihad'~\cite{rao2011love}.

This narrative carries chilling parallels to rhetoric that has catalyzed genocides in Myanmar and Sri Lanka~\cite{anwary2020interethnic}, and fueled white supremacist actions such as the Christchurch massacre~\cite{quek2019bloodbath}.
Prior work documented the wide spread use of such fear rhetoric in certain WhatsApp groups~\cite{saha2021short} or being used by Twitter influencers~\cite{dash2022insights}.

\noindent (iii) \textbf{Discrediting the Opposition}.
Lastly, we saw messages with concerted effort to discredit opposition figures, particularly leaders from the Congress party. Rahul Gandhi was a frequent target, with posts questioning his religious identity and mental acumen. There were also derogatory allegations against historical figures like former Prime Minister Jawaharlal Nehru, portraying him as a womanizer or casting doubts on his religious affiliations. These messages further accused the opposition of being anti-Hindu by alleging that they restrict Hindu festival celebrations and labeled the Congress as a `Muslim party.'


\subsubsection{Indonesia:}


\noindent (i) \textbf{Health (mis)information}. A common narrative in Indonesian WhatsApp groups mentioned alternative ways of gaining health that do not involve traditional medical help. Posts that have this narrative would usually have a self-proclaimed doctor or a set of testimonies from unidentifiable people explaining the benefits of alternative health methods. 

There were two types of viral health content that show this narrative: preventive and corrective. Preventive posts mentioned alternative ways to achieve health before illnesses occur. For example, a preventive post reminded people to clobber houses and offices to prevent dengue fever. Another example advised people to send a spiritual message about Jesus to at least 13 other groups if they do not want to experience sickness in the next two weeks.

Corrective posts, on the other hand, mentioned ways to achieve healing once illnesses have been experienced. For example, a viral post promoted the ability of raw papaya juice to increase platelet levels instantly, healing dengue fever without the need for medical help. Another example mentioned onions as the absolute medicine to self-heal from poisonous snake bites. Although the purpose of the preventive and corrective content was different, they served the same narrative: to be healthy without medical help.


\noindent (ii) \textbf{Anti-Indonesian, pro-China politics}. We found that none of the viral content in Table \ref{tab:combined_categories} were in support of Indonesia's current president, Jokowi, and his family of political elites. Multiple viral political content talked about how Gibran, Jokowi's son, should not be allowed to be a vice presidential candidate. Posts of people talking about Jokowi's ``bad policies" were prevalent. These posts were trying to convince people to go against Jokowi and his family's administration. This was highly relevant to Indonesia's 2024 presidential election, where Gibran was one of the vice presidential candidates. These anti-Indonesian politics content wished to discourage people from electing Gibran and strengthening Jokowi's political dynasty.

Interestingly, we also found plenty of Chinese political and entertainment content in our dataset. Unlike the viral political content related to Indonesian politics, none of the viral Chinese-related content contained any criticism of China. Instead, these posts often featured comparisons with other countries—most frequently the United States, which was commonly positioned as an antagonist. For example, one post mocked the U.S. for its so-called superiority complex toward China, pointing out that it has homeless citizens while China boasts advanced infrastructure. These posts consistently highlighted China’s strengths to portray it as superior not only to the U.S. but to any other country, including Indonesia.

\noindent (iii) \textbf{Advertisements}. The prevalence of advertisements in the WhatsApp conversations of Indonesian university students was at a notable 31\%, reflecting a daily inundation of promotional content ranging from academic self-development programs to job listings and Chinese products. A significant portion of these advertisements, over half, promoted academic self-development opportunities such as seminars, webinars, and internship programs, often featuring eye-catching speakers with impressive but not necessarily relevant qualifications. 

Moreover, the emotional appeal extended beyond academic advertisements to include products, exemplified by viral ads for Chinese-manufactured goods like water filters. These ads often featured dramatic demonstrations and inspirational messages, targeting the viewers' empathy and environmental concerns. This emotional strategy was a common thread linking various forms of content, including misinformation, suggesting that emotive content significantly enhances virality among this demographic. 

\subsubsection{Colombia:}

\noindent (i) \textbf{Regional Updates and Alerts}.
A significant portion of WhatsApp group activity in Colombia was dedicated to sharing regional information (21.6\%), with many groups existing solely to disseminate updates on local events. The most common types of regional updates included announcements about road closures, protests, accidents, and violent incidents. These updates often served as real-time alerts, providing crucial information to local communities.

Interestingly, this type of content was not confined to groups specifically created for regional news but was frequently shared across various group types, including family, student, and political groups, underscoring its importance to everyday communication. Beyond regional updates, these messages often included reports of missing persons/pets, or stolen vehicles, fostering a sense of communal responsibility and cooperation. Another prominent aspect involved warnings to the community about potential dangers, ranging from alerts about impending natural disasters to descriptions of new criminal tactics, such as scams or theft methods. By circulating this information, groups played an essential role in raising awareness and promoting vigilance among their members.

\noindent (ii) \textbf{Support and Discrediting Politicians}.
Political messages represented some of the most common viral content circulating in Colombia (23.8\%), with misinformation being a significant portion of it (20\%). These politically charged messages primarily focused on supporting or discrediting national-level politicians. Gustavo Petro, the current President of Colombia, emerged as the central figure in these narratives, serving as the main target of both supportive and critical messages. However, other national and regional politicians were also frequent subjects of these polarized communications.

These messages were often highly emotionally charged and utilized various formats to maximize their impact. A common format included video clips of politicians speaking or segments from news outlets overlaid with text featuring highly charged, and often satirical, commentary. Another prevalent format involved influencer-style videos, where a narrator conveyed the message, often mixed with supporting clips or visuals.
Interestingly, during the period studied, a recurring narrative emerged referencing Mexico’s recently elected president, Claudia Sheinbaum. These messages, predominantly circulated in left-leaning political groups, frequently portrayed Sheinbaum as an exemplary leader for the global left to follow and admire.

\noindent (iii) \textbf{Advertisements}.
Advertisements were very common (23.3\%), and while there were groups exclusively dedicated to sales and job opportunities, the sharing of such content was not limited to these spaces. These advertisements encompassed a broad range of topics. Event posters and agendas were particularly prevalent, showcasing a variety of activities ranging from local community gatherings to larger organizational events. Job postings were another significant category, often shared within groups specifically dedicated to job opportunities, although they also appeared in more general groups. These postings frequently included positions from both private companies and public institutions.

Additionally, course advertisements, scholarships, and entrepreneurship program calls were widely shared. These included opportunities promoted by both private and public entities, with a recurring theme of calls for alternative education programs targeting community initiatives and vulnerable populations. Another notable type of advertisement involved sales, particularly of clothing, which was often presented through catalogs. 
%
While there are parallels with the prevalence of advertisements in Indonesia, such as the focus on educational opportunities, the Colombian dataset differed by showcasing a broader emphasis on localized economic activities, including informal sales practices and regional job postings.

\noindent (iv) \textbf{Well-Being and Inspirational Content}.
An interesting similarity with South Asian contexts was the prevalence of inspirational and informational content (15.4\%), which frequently focused on personal well-being. However, in Colombia, while still common, this type of content appeared less frequently compared to India and Indonesia. Reflections on life, habit advice, and similar topics were significant examples within this category. Another recurring theme was pro-environmental information, such as messages about conservation and sustainability. 


\subsection{Misinformation}
As we can see in 
Table~\ref{tab:combined_categories},  one of the most prevalent category of viral content was misinformation across the three regions (26\% in India, 30.5\% in Indonesia, 10.5\% in Colombia). 
Notably, for all the countries, nearly half of these were in video format.
In the rest of this section, we will briefly touch upon the various misinformation narratives in the three countries.

\subsubsection{India:} The false narratives in India primarily centered on three themes: pro-BJP propaganda, anti-Congress content, and religious propaganda specifically targeting Hindus. Within these, a significant subset (44\% of the misinformation) was anti-Muslim content. This intersection of hate speech and misinformation is extremely problematic and is notably higher than rates observed in other contexts~\cite{chauchard2022circulates} or WhatsApp datasets~\cite{garimella2020images}. The virality indicates not just the prevalence but also the impact and reach of these messages. 

We also found that the misinformation was not random, but tailored to resonate with the prevailing sociopolitical sentiments. A case in point is a message circulating the `Love Jihad' myth~\cite{rao2011love}, falsely claiming Hindu women were being lured by Muslim men. This example is indicative of the dual nature of misinformation: the capability to serve as both a mirror and a molder of public opinion.
%

However, most striking is the role of mainstream media in this ecosystem. Our qualitative analysis revealed a concerning synergy between biased national media and the misinformation cycle. Television news clips favoring BJP were often shared virally in our dataset, reinforcing and legitimizing the narratives initially propagated as misinformation. 
In such a scenario, misinformation gains a veneer of credibility, amplifying its social impact and making it much more challenging to counter.

We found that misinformation in groups was not purely cognitive, but also affective; it capitalized on emotional resonance~\cite{zollo2015emotional}. This is in line with \citet{garimella2020images} who show that  during key events like elections or riots, the urgency and shock value of messages surge, often inflating simmering tensions into flashpoints of potential violence. 

\subsubsection{Indonesia:} Contrary to the Indian case, we found that in Indonesia, false narratives were mainly in health advice and religion.
Health misinformation was very dominant in our dataset, with more health-related viral content identified as false narratives (16 posts) instead of those that are not (12 posts). 
A lot of these health advice misinformation were also properly tailored to gear towards college students' lifestyles. For example, there was a post saying that a college student ate chocolate right after having instant noodles for dinner and died. 
This story has been fact checked as false by the Indonesian Communication Agency. However, the narrative was still widely forwarded and was not counter-argued by any of the group members.

Similarly, religious misinformation in our dataset was also geared towards college students' lifestyles or events. Many viral religious posts incorporated unverifiable testimonies from unnamed individuals, blending moral lessons with fictional narratives to engage audiences. Additionally, much of the misinformation in our dataset exploited group chat members' emotions—such as guilt, fear, or hope—to drive virality.


\subsubsection{Colombia:}

Misinformation in Colombia, while varied in its manifestations, can be categorized into three primary types: political misinformation (45.3\% of the messages tagged as misinformation), health-related misinformation (17.2\%), and fear-inducing alerts and conspiracy theories (31.2\%). These categories encompassed nearly all examples of misinformation found in the dataset with the remaining 6.2\% of misinformation referring to national or regional news containing claims that were verified as false.

The first category, political misinformation, was the most prevalent in the dataset, which is closely linked to the widespread presence of political propaganda in Colombia. This type of misinformation frequently involved exaggerations, inaccuracies, or outright falsehoods aimed at either supporting or discrediting political figures, parties, or ideologies. In line with the findings from India and Indonesia, many messages attacked or defended prominent national figures, perpetuating divisive narratives that contributed to the existing societal polarization. While some of this misinformation appeared to be deliberately crafted, much of it was spread by individuals who genuinely seem to believe the narratives, underscoring the blurred line between disinformation and misinformation in this context.

The second major type of misinformation involved health advice and medical information that lacks scientific backing or directly contradicts established medical knowledge. Messages in this category wee shared in various formats, including videos, images, and plain text. Perhaps the most prevalent example was anti-vaccine information, which frequently overlaps with conspiracy theories.  Such health-related misinformation  ranged from claims about viral infections and preventative care to advice on managing age-related conditions. The latter particularly appeals to older populations, who are known to be more susceptible to misinformation \cite{brashiere2020}. 

The third category included fear-inducing alerts and conspiracy theories, often featuring sensational claims to spread panic, caution, or distrust. For example, a long-running false narrative in Latin America since 2016 claims escaped convicts are abducting children, often accompanied by photos of alleged criminals. This message frequently reappeared in the dataset, sometimes in altered forms, even in distant groups on the same day. Other fear-based content included exaggerated warnings about crime or personal safety. The persistence and wide reach of such messages suggest they may be amplified by individuals or unknown actors, although their exact motives remain unclear.


The Colombian case mirrors India in the high prevalence of political misinformation and aligns with Indonesia in the prominence of health-related misinformation. However, unlike India and Indonesia, religious misinformation was very rare in Colombia. Only three posts in the dataset could be classified as religious propaganda susceptible to fact-checking. This reveals both unique regional nuances and important similarities across vastly different contexts.

\subsection{Prevalence of entertainment content}

\citet{arora2016bottom} suggests that ``people go online to romance, game, be entertained, consume media,
view pornography, and share their personal thoughts and feelings.''
We found this to be certainly the case on WhatsApp too.
Our analysis revealed that entertainment is a predominant category of content consumed. We found  a diverse range of entertainment materials, encompassing humor clips, non-political commentary, educational pieces on well-being, information about government programs, sports, beauty, art, gossip,
and more. Such messages along with religious content, political satire, and ``good morning" messages constituted over a third of the virally disseminated materials~\cite{wsjInternetFilling} in the three contexts.

Interestingly, contrary to findings from other studies involving similar demographics~\cite{rangaswamy2016mobile,vashistha_sangeet_2015} entertainment was not the most prevalent type of viral content. Misinformation and news dominated the content shared in South-Asian context, while Political propaganda (23.8\%) and regional information (21.6\%) dominated in the Colombian context. This is surprising given the demographic profiles--young men in India or students in Indonesia. One might have expected a greater prevalence of entertainment content, reflecting their age and interests.

This suggests that WhatsApp might play a different role compared to other social media platforms for these demographics, serving less as a space for leisure and more as a medium for information dissemination, including misinformation. It raises questions about the influence of WhatsApp’s private and closed nature on content dynamics, where the spread of misinformation and news may be amplified in a way that differs from public platforms.
Larger, representative quantitative studies are needed to explore these dynamics further and to understand why entertainment takes a backseat to more serious content in these settings.


\subsection{Fact checking and its impact}

Our analysis highlights a critical and concerning observation: efforts to correct or fact-check misinformation within group chats were extremely rare. Among the 158 instances of misinformation identified in India, 53 in Indonesia, and 64 in Colombia, only one instance in India and Indonesia involved active correction or rebuttal within the group. This reveals a troubling pattern of widespread passive acceptance of misinformation across diverse contexts.
This also extends beyond just the misinformation content, to other types such as hate speech, and propaganda. Surprisingly, we found that no form of fact-checking ---be it images, videos, text, or links--- was present in all three regions. This scenario is paradoxical, especially since much of the misinformation circulating in these groups had already been debunked by mainstream fact-checking agencies. 

This implies one of two scenarios: either the fact-checks are failing to penetrate the circles in which misinformation circulates, or they are ineffectual in deterring users from sharing previously debunked information. Our data leans towards the former hypothesis. We found the resilience of debunked content to be remarkable -- 60 out of 158 misinformation posts in India, 17 out of 53 in Indonesia, and 20 out of 64 in Colombia had already been fact-checked. We noted several instances where identical pieces of misinformation related to health or fear-inducing alerts have been systematically fact-checked annually by reputable agencies since 2016. Yet, these debunked claims continued to disseminate widely on WhatsApp, reflecting the lack of effectiveness of current fact-checking strategies in reaching the target audiences.  These results align with the study by~\citet{seelam_fact-checks_2024}, which highlights significant challenges faced by fact-checking organizations in India when it comes to reaching audiences.

Apart from top-down fact checking, we also looked for instances of within group, community driven fact-checking~\cite{malhotra2023misinformation}. These are instances where a member of the group corrects misinformation posts. We found that there was only one instance of such correction in India and Indonesia (none in Colombia) out of the hundreds of instances of misinformation posted. Previous research in this space~\cite{ng2022self,varanasi2022accost,shahid_one_2024} shows that such corrections do not happen often due to social pressures and power dynamics attached with correcting people in front other group members, which we think could also be at play in the groups we study.


The staggering ineffectiveness of current fact-checking mechanisms calls for a radical rethinking of our approach. Given the trust dynamics and the persistence of debunked content, a more integrated and proactive fact-checking mechanism involving platform operators, community leaders, and policy stakeholders seems imperative.

\section{Discussion}
\label{sec:discussion}
The paper presents a mixed-method analysis of viral content spreading in 164 Indian, 97 Indonesian and 140 Colombian private WhatsApp groups. We collected the WhatsApp groups data directly from villages in rural India, university students among Indonesian universities, and Colombian participants nationwide. This is the first study to examine viral content in private WhatsApp groups at such a scale in the Global South, giving insights at everyday consumption both in terms of the types of groups users consume information from as well as the viral content that spreads in these groups.
Though the samples from the three regions are completely different, data reveals a surprising amount of similarity in terms of the types of groups people are a part of, and the types of viral content they receive.

\subsection{Content Prone to \textit{Virality} and Misinformation}
In both India and Indonesia, we found that both religious and political content were more likely to be viral, often containing misinformation and lacking alternative viewpoints. In India, anti-Muslim content freely propagated. These narratives appear to be orchestrated to amplify specific, often troubling themes—such as hate speech targeting Muslims—thus warranting serious consideration of their societal and national ramifications. Such a constant deluge of divisive content raises concerns about its long-term psychological impact, particularly in fostering a climate of hostility towards minorities. While the specific media consumption habits of Muslim individuals within this context remain unknown, the overall trend is troubling. Invoking Anderson et al.’s concept of the ``saffronization of the public sphere," \cite{anderson2020} the findings highlight an unsettling trajectory potentially leading to offline violence, as observed in situations like Sri Lanka and Myanmar. This study offers empirically grounded alerts, demanding immediate policy interventions to curb this escalating issue before large scale violent clashes between communities threaten the fabric of the country.

In Indonesia, viral religious content were indeed \textit{tamer} compared to those in India. Most of them did not convey hate speech towards other religions and were mainly posts that may possibly help group members to \textit{spiritually grow}. However, these same content included many unverifiable consequences and also ``divine punishments'' to those who did not forward, often within a specific deadline and to a specific number of people and followed the doctrines that were written in these religious posts. Instead of instilling hatred towards other groups of people, like in India, viral Indonesian religious content were created to instill fear and emotions into people who consume the content. This fact itself is troubling since emotions increase the belief in fake news \cite{martel2020}. The concerns following this emotional approach to religious content is that it will lead to sharing due to fear and online indirect coercion instead of personal desire. Research has also found that people would feel obliged to share misinformation if it means helping others avoid misfortune \cite{varanasi2022accost}. It is possible, then, that Indonesian religious content will only worsen the spread of misinformation, as readers fear the possibilities of experiencing and causing misfortune. This study once again alerts the need of new intervention and educational methods to help WhatsApp users in the Global South to be able to prioritize recognizing fake information instead of feeding the fear and emotions originating from forwarded religious content.

Instead of having more local political content, our findings showed that Indonesians tend to share more of Chinese politics. In contrast to the negativity surrounding Indonesian politics and the Jokowi administration, viral Chinese political content were all positive in nature. China was portrayed as the most powerful nation globally, often being compared to other countries such as the United States, Japan, and even Indonesia. Once again, the lack of alternate views on Chinese politics can instill narrow and radical judgments among Indonesians, which is an urgent matter in itself. If viral content such as in our findings continue on, it is possible that there will be a decrease in nationalism and patriotism among Indonesians, replaced instead by the desire to move abroad to countries promoted to have better governmental administrations, even when in reality may not always be the case.

Colombia presents a unique blend of challenges when it comes to content prone to virality and misinformation. Regional updates dominate the landscape, reflecting the localized and community-driven nature of WhatsApp group activity. However, political misinformation stands out as the most prevalent, often characterized by emotionally charged narratives designed to support or discredit key political figures such as Gustavo Petro. This type of content echoes the divisiveness seen in India’s political misinformation, amplifying polarization and entrenching societal divides. Another critical category involves health misinformation, which mirrors patterns observed in Indonesia, with a focus on anti-vaccine narratives and and unsupported health advice. While religious misinformation is notably rare in Colombia, fear-based content remains a significant concern, including persistent and exaggerated warnings about criminal activities. These findings underline the necessity of targeted interventions in Colombia to address the distinct dynamics of misinformation, from fostering digital literacy to enhancing fact-checking efforts, while considering the country’s unique sociopolitical and cultural context.

\subsection{The Struggle to Combat Misinformation}

Our research reveals a disconcerting reality: fact-checking appears to be an exercise in futility within these WhatsApp groups. 
Misinformation was almost never debunked despite being fact-checked, highlighting a deep-seated issue with information veracity in these settings. Moreover, there is a pervasive lack of awareness --or perhaps interest-- about the concept of fact-checking among group members \cite{seelam_fact-checks_2024}. The reactive nature of current fact-checking methods is exacerbated by WhatsApp’s end-to-end encryption, making it impossible to proactively counter misinformation at the source. Even narratives that have been discredited nearly a decade ago continue to circulate virally, undeterred by the current fact-checking efforts.

The current model for fact-checking on WhatsApp relies on tiplines, which operate on an opt-in, reactive basis~\cite{whatsappIFCNFactchecking,kazemi2022research}. This is fundamentally flawed for several reasons. First, mass adoption is lacking. Second, and perhaps more crucially, trust in the WhatsApp ecosystem typically arises through personal networks, creating a barrier to external information, including fact-checks.

This situation raises critical questions about the effectiveness of current methods and the urgent need for innovative solutions that can adapt to the unique challenges posed by encrypted platforms. It implies that the battle against misinformation may need to shift from simply debunking false narratives to fundamentally altering the ecosystem that allows these narratives to thrive unchecked.

The real challenge, then, is not just writing or updating fact-checks, but ensuring they reach the appropriate audiences --- a feat that current methods are failing at. In this context, a more grassroots approach may be essential. Leveraging community-driven fact-checking and deliberation models~\cite{agarwal_conversational_2024}, which are designed to be participatory and decentralized, might be one of the few sustainable ways to combat this issue. Such a system could better capitalize on the inherent trust and social ties within WhatsApp groups, creating a more resilient and adaptable mechanism for information verification.

Overall, the study advances the field of social computing by highlighting how technology which is currently understudied in the ICWSM community intersects with social, cultural, and political factors. It invites the ICWSM community to explore new avenues for research and practice that can lead to more effective and equitable technology design, ultimately contributing to the betterment of global digital societies.

\subsection{Limitations}
Our research comes with several limitations that warrant discussion while interpreting the results:

\noindent\textbf{Sample size and convenience sampling}: While our dataset presents a disturbing prevalence of politically charged and misleading narratives, it is important to remember that the data originates from a relatively small, right-leaning convenience samples in India, Indonesia, and Colombia. Additionally, to protect user privacy, we only chose groups with a certain size and activity, thus missing out personal conversations where misinformation could be shared. Consequently, the scope for generalization remains limited. However, the presence of such troubling content in even a small sample raises urgent questions about the extent and depth of the issue in larger, more diverse populations.

\noindent\textbf{Observational nature}: The study is essentially observational, limiting our ability to ascertain the intent behind the forwarding of misinformation or the belief in the misinformation. While our analysis focuses on content exposure rather than spread dynamics, the data doesn't reveal whether individuals forwarded messages knowingly or unknowingly. This limits our understanding of user motivations and the psychological underpinnings of the dissemination process.

\noindent\textbf{Prevalence vs. Exposure}: 
In our analysis, it is vital to clarify that the proportion of misinformation among viral messages does not directly translate into the same proportion of misinformation in the total messages each user encounters. However, this distinction should not minimize the severity of our findings. While misinformation may not dominate an individual's daily message feed, the fact that debunked misinformation is still reaching a viral status---especially when laden with hate speech---is deeply concerning. This recurring cycle of misinformation, even after public debunking, points towards a systemic issue that extends beyond the scope of individual exposure and speaks to a larger, more pervasive problem.


\noindent\textbf{Geographic and cultural context}: While the study is focused on a specific contexts --- e.g., rural India or Indonesian students ---  one might argue that its findings have limited applicability beyond this particular demographic. However, we posit that our methodology and findings have broader implications. 
The mechanics of misinformation and political propaganda can be similar across different contexts, making our research a valuable case study for understanding the dynamics on end-to-end encrypted platforms globally.
The findings, particularly the observed prevalence of politically charged content and misinformation, contribute to a deeper understanding of how WhatsApp is used in such a context. This is especially relevant given the limited research in this specific area. Some of the findings, such as the prevalence of hyper partisan messages, is interesting even in this limited context because it shows (though not fully) the range and reach of political parties and their WhatsApp information ecosystem.

These limitations, far from undermining the study, serve as important qualifiers that provide direction for future research. They raise compelling questions about the true extent of misinformation spread and political propaganda on encrypted platforms, while also pushing for nuanced approaches that account for local context and individual behavior. 

\subsection{Future Work}
As the first analysis of its kind, this study establishes a foundation for further investigations into the dynamics of content dissemination on WhatsApp across diverse global contexts. While our focus on widely circulated messages marked as ``forwarded many times'' captures prominent themes and attitudes, future research should also expand beyond viral content. There is a need to develop scalable yet ethically sound models of data collection that can capture a broader spectrum of content, enabling researchers to assess the prevalence and characteristics of misinformation more comprehensively~\cite{sehat2021ethical}.

Future work could combine quantitative and qualitative methodologies to provide deeper insights into the impacts of content shared on WhatsApp. Integrating qualitative interviews or focus groups with quantitative analyses can reveal how users perceive, interpret, and respond to the content they encounter, thus enriching our understanding of the social implications of misinformation and other types of viral messages. Such combined methods will be particularly valuable in contexts like those studied here---India, Indonesia, and Colombia---where cultural nuances significantly shape digital communication practices.

\section*{Acknowledgements}
Kiran Garimella is supported by a National Science Foundation grant ``Towards a Privacy-Preserving Framework for Research on Private, Encrypted Social Networks'' (ID \#2318843) and a Knight foundation grant.
Aditya Vashistha thanks the Mario Einaudi Center for International Studies and Cornell Atkinson Center for Sustainability for supporting this work.


\bibliographystyle{aaa25}
\bibliography{biblio}

\clearpage

\appendix

\section{Appendix}

\subsection{Definitions of group categories (India)}

We manually classified all the 164 groups from India into 10 categories. Table~\ref{tab:groups_annotation_appendix} shows the definitions of the group categories.

\begin{table*}[]
\caption{Descriptions of the group categories.}
\label{tab:groups_annotation_appendix}
\centering
\begin{tabular}{ll}
\hline
\textbf{Category} & \textbf{Description} \\
\hline
\begin{tabular}[c]{@{}l@{}}Village \end{tabular} & \begin{tabular}[c]{@{}l@{}}Has the village name in it  and is apolitical\end{tabular} \\
\hline
\begin{tabular}[c]{@{}l@{}}Caste\end{tabular} & \begin{tabular}[c]{@{}l@{}}Has the name of any caste\end{tabular}\\
\hline
\begin{tabular}[c]{@{}l@{}}Religious\end{tabular} & \begin{tabular}[c]{@{}l@{}}Has the name of any religion, a god or \\ symbol of a religion \end{tabular}\\
\hline
\begin{tabular}[c]{@{}c@{}}Hindutva\end{tabular} & \begin{tabular}[c]{@{}l@{}}Has the BJP, a hindutva idealogue, or a \\ hindutva group\end{tabular}\\
\hline
\begin{tabular}[c]{@{}l@{}}Activism\end{tabular} & \begin{tabular}[c]{@{}l@{}}Is non-political and demands any rights\end{tabular}\\
\hline
\begin{tabular}[c]{@{}l@{}}Regional\end{tabular} & \begin{tabular}[c]{@{}l@{}}Has the name of any village, town,\\ city, district of Jharkhand and not \\ any of the five categories listed above\end{tabular}\\
\hline
\begin{tabular}[c]{@{}l@{}}Friends\end{tabular} & \begin{tabular}[c]{@{}l@{}}Refers to a friend group\end{tabular}\\
\hline
\begin{tabular}[c]{@{}l@{}}News\end{tabular} & \begin{tabular}[c]{@{}l@{}}Primarily focused on sharing and discussing current events or updates from official, unofficial, or community sources. \end{tabular}\\
\hline

\begin{tabular}[c]{@{}l@{}}Others\end{tabular} & \begin{tabular}[c]{@{}l@{}}Can not be categorized into any \\ of the six groups listed above \\
e.g. fun group, hobbies, etc.\end{tabular}\\
\hline
\end{tabular}
\end{table*}

\subsection{Definitions and examples of content categories (India)}

\begin{enumerate}

\item Misinformation: Incorrect or misleading information. Since the annotator was a trained fact-checker, any claim which we found to be false by fact-checking them was tagged as misinformation. 
There were contents which could not be fact-checked, e.g. opinions. These are not included in this category.

\item Information/Inspirational videos/Commentary/Amusement videos/Religious harmony/Educational messages: 
A catch-all category containing videos providing information, inspirational videos, providing commentary/opinion about certain (non political) issues, educating people, or promoting religious harmony.

\item Religious propaganda to influence Hindus: Messages targeted to make Hindus feel that their rights are being taken, they are discriminated, deprived of their religion, fear mongering about the extinction of Hindus/Hinduism, etc.

\item Hate against Muslims: Characterized by or expressing hostility or discrimination toward Muslims or the Islamic faith, including content that clearly and deliberately incites hatred against Muslims.
Examples include: asking Hindus to unite, Muslims rulers being looters and thus justifying attrocities against Muslims, Muslims being traitors/Pakistan supporters, etc.

\item Pro-BJP political propaganda: Information--facts, arguments, rumours, half-truths, or lies--to influence public opinion in favor of the BJP.
Includes any post that endorses BJP, Uttar Pradesh CM Yogi Adityanath, Prime Minister Modi or defends BJP, or other BJP leaders.

\item Anti-Congress political propaganda: Content targeted towards the Congress party or any of its leaders Nehru, Sonia Gandhi, Rajiv Gandhi or Rahul Gandhi, etc.

\item Regional information: Regional news or are related to local demand of Jharkhand. The messages in the regional news were mostly related to demands for jobs.

\item Religious: Any content which is not political and is not right-wing or left wing but contains mentions of god.

\item Good morning messages: Messages wishing people good morning.

\item Political or religious sarcasm/satire: 
 Spoof, Humor, Sarcasm to advance political or religious propaganda.

\item Political opinion not to benefit any political party:
Political opinion not to endorse or malign any particular party.

\item Health misinformation: Any health claim that lacks evidence or a claim that goes against current evidence.

\item Anti-BJP propaganda: Content targeted against BJP and its leaders.

\end{enumerate}

\subsection{Definitions and Examples of Content Categories (Indonesia)}
\begin{enumerate}
\item Local information: Viral posts containing information with specific regional boundaries mentioned that are the size of a city or smaller.
\item Advertising: Viral posts promoting events or products that are accompanied by further contact methods for readers to RSVP or text sellers.
\item Misinformation: Incorrect or misleading information. Both posts that are verified by the Indonesian Communication Agency as 'HOAX' and those that are unverifiable due to the lack of related information outside of the posts are included here.
\item Inspirational Message: Viral posts that promote non-health behavioral changes in order to achieve specific non-health goals, usually in the shape of advice.
\item National Information: Viral posts containing information with specific regional boundaries mentioned that are the size of a province or bigger, but still within the scope of Indonesia
\item Entertainment/Humor/Sarcasm: Posts that are shared to make people happy or laugh. This includes beauty of nature, cultural posts, people doing funny acts, cartoons, etc.
\item International Information: Posts containing information from outside Indonesia
\item Religious: Posts that explicitly mentioned the doctrine or verses of at least one of the 6 official Indonesian religions: Islam, Protestantism, Catholicism, Buddhism, Hinduism, Confucianism 
\item Health Advice: Posts that include health content
\item Chinese Content: Posts that include content about anything Chinese: Chinese products, Chinese politics, Chinese nature, Chinese culture, Chinese technology, posts that are written or spoken in Chinese.
\item Religious Iconography: Posts that explicitly mentioned the characters/people of at least one of the 6 official Indonesian religions: Islam, Protestantism, Catholicism, Buddhism, Hinduism, Confucianism 
\item Political: Anything explicitly concerning the government no matter the scope: Indonesia, international, the world.
\item Propaganda: Anything political that aims to change behavior or opinion by promoting or depromoting a specific government/party/person
\item Hateful Content: Posts that show hostility or discrimination directly towards a specific country/group/person
\item Partisan: Political posts that show obvious support or dislike towards a specific political party/character.
\item Good Morning Message: Messages wishing people good morning
\end{enumerate}

\begin{table*}[]
\caption{Descriptions of the group categories.}
\label{tab:groups_Colombia_appendix}
\centering
\begin{tabular}{ll}
\hline
\textbf{Category} & \textbf{Description} \\
\hline
\begin{tabular}[c]{@{}l@{}}Study-related \end{tabular} & \begin{tabular}[c]{@{}l@{}}Has the name of a school, university, or class in it.\end{tabular} \\
\hline
\begin{tabular}[c]{@{}l@{}}Work related\end{tabular} & \begin{tabular}[c]{@{}l@{}}Has the name of a company or is focused on professional activities.\end{tabular}\\
\hline
\begin{tabular}[c]{@{}l@{}}Sales\end{tabular} & \begin{tabular}[c]{@{}l@{}}Includes sales and commerce groups with these words in their names. \end{tabular}\\
\hline
\begin{tabular}[c]{@{}c@{}}Jobs\end{tabular} & \begin{tabular}[c]{@{}l@{}}Includes job-seeking groups and other kind of opportunities.\end{tabular}\\
\hline
\begin{tabular}[c]{@{}l@{}}Family\end{tabular} & \begin{tabular}[c]{@{}l@{}}Refers to family groups, often including the word ``family" or a surname.\end{tabular}\\
\hline
\begin{tabular}[c]{@{}l@{}}Friends\end{tabular} & \begin{tabular}[c]{@{}l@{}}Refers to a friend group.\end{tabular}\\
\hline
\begin{tabular}[c]{@{}l@{}}Regional \& Local \end{tabular} & \begin{tabular}[c]{@{}l@{}}Has the name of a neighborhood or city and is usually focused on local news and relevant information. \end{tabular}\\
\hline
\begin{tabular}[c]{@{}l@{}}Hobbies and Activities\end{tabular} & \begin{tabular}[c]{@{}l@{}}Refers to groups with names of sports teams or specific activities like cycling, traveling, etc. \end{tabular}\\
\hline
\begin{tabular}[c]{@{}l@{}}Religious\end{tabular} & \begin{tabular}[c]{@{}l@{}}Has the name of a religion or religious community in it.\end{tabular}\\
\hline
\begin{tabular}[c]{@{}l@{}}Others\end{tabular} & \begin{tabular}[c]{@{}l@{}}Can not be categorized into any of the six groups listed above.\end{tabular}\\
\hline
\end{tabular}
\end{table*}

\subsection{Definitions of group categories (Colombia)}

We manually classified all the 140 groups from Colombia into 12 categories. In the main document, the category ``Hobbies and activities" is reported within the category ``Other" to be consistent with the Indonesian classification. Table~\ref{tab:groups_Colombia_appendix} shows the definitions of the group categories.

\subsection{Definitions and examples of content categories (Colombia)}

\begin{enumerate}

\item \textbf{Misinformation:} Incorrect or misleading information. The main annotator performed a first classification and search, where some suspicious elements were already fact-checked. In addition, we received help from a trained fact-checker, who gave a second review to the suspicious contents.
There were contents which could not be fact-checked, e.g. opinions or unverifiable regional information, such as accidents, crimes or missing persons. These are not included in this category.

\item \textbf{Information/Inspirational Content/Reflections/Educational Messages:} A catch-all category containing videos providing information, inspirational content, commentary or opinions about certain (non-political) issues, educating people, or promoting well-being.

\item \textbf{Religious Propaganda:} A very rare category in Colombia, it was limited to three videos showing Christian-evangelical politicians opposing a bill in Congress to prohibit conversion therapies, basing their opposition on religious arguments.

\item \textbf{Political Propaganda:} Messages with overt political tones, representing various sides of the ideological spectrum. These messages often defend political positions, promote agendas, or support/attack prominent political figures and parties. They are crafted to elicit strong emotional responses, leveraging polarizing narratives to influence opinions and foster ideological alignment.

\item \textbf{Religious Content:} Content such as prayers and reflections that refer to a deity or religion.

\item \textbf{Regional Information:} This category includes news and updates about current events at the local or regional level. Examples include reports of missing persons, pets, or vehicles, as well as accidents and murders. It also encompasses information about road closures, regional documentaries, and local government announcements, such as service suspensions or road blockages. These messages often serve to inform and engage communities on issues directly affecting their immediate surroundings.

\item \textbf{Conspiracy Theory:} This category includes conspiracy narratives such as anti-vaccine rhetoric and related theories, like those claiming Bill Gates’ involvement in population control or microchip implantation. These messages are often intertwined with misinformation, reinforcing their appeal and spreading falsehoods widely across groups.

\item \textbf{Advertisement:} Content that promotes events, job postings, scholarships, entrepreneurship and educational opportunities, or other types of advertisements.

\item \textbf{Sales and Commerce:} Content promoting products or services, often accompanied by further contact methods for purchasing.

\item \textbf{National/International News:} News content at the national or international level, frequently presented as clips sourced from mainstream media. These messages aim to inform about significant events and developments but occasionally circulate without verification, blending legitimate reporting with potential misinformation.

\item \textbf{Entertainment/Humor/Sarcasm:} Posts shared to make people happy or laugh. This includes images or videos about the beauty of nature, cultural content, funny acts, cartoons, etc.

\item \textbf{Health Information:} Medical information or health advice.

\item \textbf{Good Morning Messages:} Messages wishing people good morning as well as congratulations on holidays such as Mother’s Day or Father’s Day.

\item \textbf{Other:} Content that does not fit into any other category.

\item \textbf{AI Generated:} Content clearly generated by AI, usually consisting of simple images or videos.

\end{enumerate}

\end{document}